\documentclass{elsart}

\usepackage{slashed}
\usepackage{graphicx}

\newcommand{\be}{\begin{equation}}
\newcommand{\ee}{\end{equation}}
\newcommand{\bea}{\begin{eqnarray}}
\newcommand{\eea}{\end{eqnarray}}
\newcommand{\beal}{\begin{align}}
\newcommand{\eal}{\end{align}}
\newcommand{\bespl}{\begin{split}}
\newcommand{\espl}{\end{split}}

\newcommand{\nsl}{\kern 0.2 em n\kern -0.50em /}
\newcommand{\ksl}{\kern 0.2 em k\kern -0.45em /}
\newcommand{\psl}{\kern 0.2 em p\kern -0.50em /}
\newcommand{\Nsl}{\kern 0.2 em N\kern -0.50em /}
\newcommand{\ssl}{\kern 0.2 em s\kern -0.50em /}
\newcommand{\pbsl}{\kern 0.2 em \bar{p}\kern -0.50em /}
\newcommand{\sbsl}{\kern 0.2 em \bar{s}\kern -0.50em /}
\newcommand{\kbsl}{\kern 0.2 em \bar{k}\kern -0.50em /}
\newcommand{\nbsl}{\kern 0.2 em \bar{n}\kern -0.50em /}
\newcommand{\Nbsl}{\kern 0.2 em \bar{N}\kern -0.50em /}
\newcommand{\Pslash}{\kern 0.2 em P\kern -0.50em /}
\newcommand{\Rslash}{\kern 0.2 em R\kern -0.50em /}

\begin{document}

\begin{frontmatter}

\title{
Suppression of contributions from large-parton-number Fok states to 
T-odd distribution functions in Drell-Yan scattering involving   
small$-x$ annihilating quark and antiquark. 
} 

\author{A.~Bianconi}
\address{Dipartimento di Chimica e Fisica per l'Ingegneria e per i 
Materiali, Universit\`a di Brescia, I-25123 Brescia, Italy, and\\
Istituto Nazionale di Fisica Nucleare, Sezione di Pavia, I-27100 Pavia, 
Italy}
\ead{bianconi@bs.infn.it}

\begin{abstract}
T-odd distributions like Sivers and Boer-Mulders functions 
are normally modeled using few-body models.  
In the present work I want to follow the completely opposite 
point of view, and study a high-energy proton-proton Drell-Yan 
where both 
the annihilating partons are wee, and one (at least) 
comes from a high-order Fok state of the parent proton, i.e. 
a state where a large number $N$ of partons is present. 
I show that 
rescattering between active and spectator 
partons is modified, with suppression of those terms that are needed to 
build T-odd parton distributions. 
\end{abstract} 


\end{frontmatter}

\maketitle

\section{Introduction}

The measurement of the $cos(2\phi)$-asymmetry associated with the 
violation of the Lam-Tung rule\cite{LamTung78} 
in Drell-Yan dilepton production 
by the 
collaboration E866\cite{E866} 
poses the problem of the behavior 
of this asymmetry when the longitudinal fraction of 
both the annihilating partons is small. 
An evidently nonzero asymmetry was 
systematically measured 
in fixed-target pion-nucleus Drell-Yan 
experiments\cite{fixedtarget},
where the 
value of the product $x x'$ $\approx$ $Q^2/s$ 
($Q^2$ is the dilepton mass, $s$ the collision c.m. squared 
energy, $x$ and $x'$ the  
longitudinal fractions of the annihilating partons) always 
implied that one at least of the annihilating 
partons was a valence (anti)quark. 
In proton-proton Drell-Yan, sea antiquarks are necessarily involved, 
and the much larger c.m. energy of E866\cite{E866} led statistically to 
a much smaller value of $x x'$.  
E866 measured zero-compatible values of the asymmetry, 
with the possible exception of the largest$-x$ point. 

According with the modern way of seeing this asymmetry, within a 
leading-twist factorization formalism\cite{CollinsSoperSterman} it is 
proportional to the convolution of two T-odd TMDF
(transverse momentum dependent distribution functions), the 
Boer-Mulders\cite{BoerMulders98} functions. Models for the 
Boer-Mulders function, for another T-odd TMDF 
(the Sivers function\cite{Sivers}), and for the related 
asymmetries, have been studied by many authors 
(see 
e.g. \cite{EfremovTeryaev82,QiuSterman91, 
BrodskyHwangSchmidt02,Collins02,JiYuan02,BoerBrodskyHwang03, 
GambergGoldsteinOganessyan03,Yuan03,
Pobylitsa03,BacchettaSchaeferYang04,Burkardt04,LuMa04,Bomhof04}
 \cite{Drago05,GoekeMeissnerMetzSchlegel06,
JQVY06, Entropy3,BacchettaContiRadici08,
Courtoy08,AB_JPG1,matches08,Vento09,RatcliffeTeryaev09,
PasquiniYuan10}). 
The phenomenological models normally implement schemes where 
a hadron is composed by 2 or 3 constituents. This puts some 
limitations on the understanding of the small$-x$ properties of T-odd 
TMDF and related asymmetries (see however the recent 
\cite{YuanXiao10}). 

At the lowest PQCD order, 
the Drell-Yan cross section is proportional to the imaginary part 
of the amplitude $G_0$ of fig.1. Fig.2 shows the amplitude 
$G_1$ including part of the $O(\alpha_s)$ corrections 
(for the full structure of the Drell-Yan cross section, see 
\cite{ArnoldMetzSchlegel09} and \cite{BaroneDragoRatcliffe}). 
Not to violate fundamental principles, T-odd TMDF 
are necessarily associated with the 
presence of rescattering\cite{BrodskyHwangSchmidt02} 
in the Drell-Yan process. 
More precisely, with the interference 
between rescattering and no-rescattering terms. 
In a leading twist factorization scheme, 
a model for obtaining a nonzero T-odd TMDF for the parton 
coming from hadron ``2'' needs including at least fig.1, fig.2 
and its right-side specular.  

\begin{figure}[ht]
\centering
\includegraphics[width=9cm]{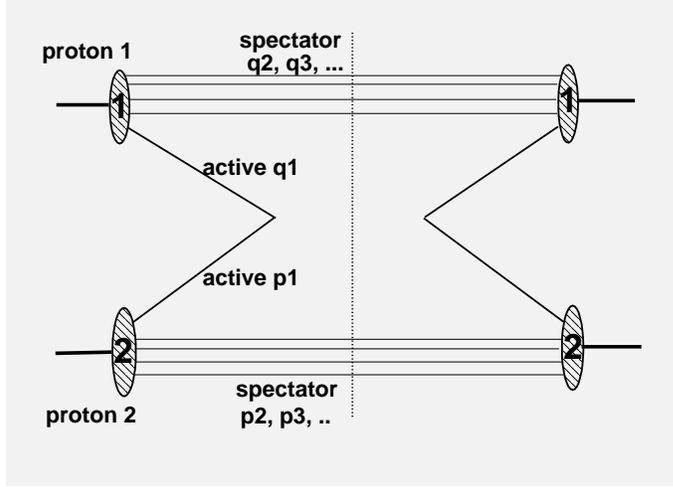}
\caption{Cut-diagram for the Drell-Yan parton-model cross 
section. The relevant particles for this work are $q_1$ (the active 
quark or antiquark from proton 1) and $p_2$, $p_3$, ... (the spectator partons 
from proton 2). For simplicity, the virtual photon and the leptons are not 
drawn. 
\label{fig1}}
\end{figure}

\begin{figure}[ht]
\centering
\includegraphics[width=9cm]{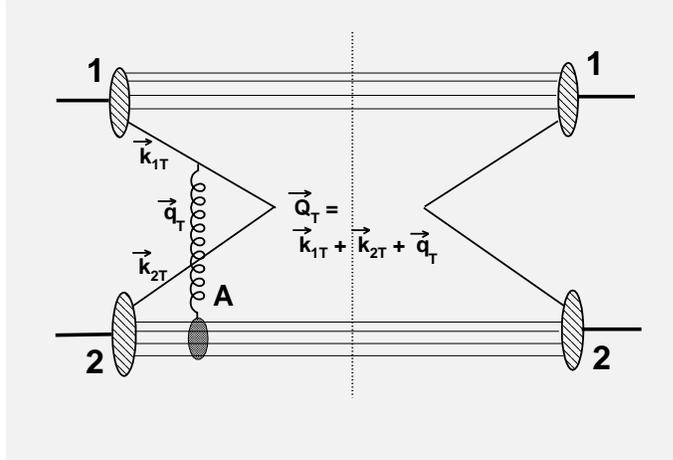}
\caption{Drell-Yan cross section including the 
active$_1-$spectator$_2$ $O(\alpha_s)$ 
rescattering amplitude $A$, with detailed 
kinematics. $\vec Q_T$ is the transverse momentum of the virtual 
photon, $\vec q_T$ the one transported by the rescattering boson, 
$\vec k_{1T}$ and $\vec k_{2T}$ those from the initial state of the 
active partons. 
\label{fig2}}
\end{figure}

Not all the rescattering terms are responsible for T-odd 
effects. A change $\Delta L$ $=$ 1 in the orbital angular momentum 
of the active parton is required. 
I will name ``spin-orbit terms'' those that may 
produce this (without violating fundamental laws), and 
``scalar terms'' those that cannot. 
I expect spin-orbit rescattering terms to become ineffective 
when states with a large number $N$ of partons, are involved. 
The qualitative arguments are: 

(i) A large $N$ state is associated with an increasing transverse 
radius $R_N$ for the individual parton distribution. 

(ii) The set of the rescattering spectator partons forms an 
overall color triplet, i.e. its total color charge does not 
depend on $N$. The effectiveness of triplet-triplet 
$O(\alpha_s)-$interactions is decreased by 
color-charge form factors when the partons 
spread over a broad region.  

(iii) This regards scalar and spin-orbit terms, but the latter 
are proportional to the transverse gradient in impact parameter 
space, so they are much more sensitive to the shape of the color-charge 
distribution.

Of course other effects may suppress spin-dependent TMDF at 
small $x$. Here I want to focus on the role of rescattering. 

\subsection{Small$-x$ limit and general assumptions} 

In the present work, I assume a large but $finite$ and $fixed$ value 
of the c.m. collision energy $s$, and 
consider a Drell-Yan process 
involving two protons where the two conditions are satisfied: 

(1) The active quark and antiquark have both longitudinal fraction 
$x$ $<$ $E_o/\sqrt{s}$ with $E_o$ $\sim$ 1 GeV. 

(2) One of the two protons (proton ``2'' in fig.2) 
is in a high-order Fok state: the number $N$ of its 
partons satisfies $N$ $>>$ 3. 

Condition (1) for both the annihilating quark and antiquark 
defines a ``small$-x$ kinematic framework'', in the sense that 

(i) although $s$ is possibly large, $Q^2$ $\approx$ 
$xx's$ is at most semi-hard. The active quark and 
antiquark momenta have fixed upper scale $\sim$ 1 GeV. 
A more extensive description of the peculiarities of small$-x$ 
physics in hard scattering may be found e.g. in \cite{kolya02}. 
In practice point (i) means that the conditions for an 
unambiguous separation of leading twist terms are not present
(see also \cite{YuanXiao10}). 
For these reasons here I will $not$ exploit 
the ordinary factorization ideas, but will anyway consider 
factorization as the scheme according to which final results 
of an experiment will be rewritten. 

(ii) longitudinal degrees of freedom disappear from the problem, 
since the active (anti)quark is not able to probe target 
structure details within (1 GeV)$^{-1}$ (a length that is 
$E/M$ times longer than the UR-contracted target thickness). 
This means that the rescattering problem is 2-dimensional. 

I will here rely on some qualitative results of the old 
multiperipheral model for hadronic collisions 
involving large$-N$ Fok states. It was first proposed in 
\cite{multiperipheral1}, and later developed and branched by a 
countless 
number of authors (for a review of the basic ideas, see 
\cite{multiperipheral2}). 
It organizes the (interacting) hadrons in chains 
of $N$ partons whose impact-parameter wavefunction 
is random-walk structured: 
$\psi(\vec b_1, \vec b_2, ...)$ $\approx$ 
$\prod \psi(\vec b_i - \vec b_{i-1})$. 
So the overall average squared transverse radius satisfies 
$R_N^2$ $\propto$ $N$. 

I will only include $O(\alpha_s)$ processes. Since at this order 
the commutativity problem is not present, I will 
treat color charges as abelian QED-like charges, 
assuming that a Drell-Yan event splits a hadron into 
an active (anti)quark with color charge $-1$ and a spectator 
set with color charge $+1$.


\section{N-spectator rescattering}

At large $E/M$ the 4 independent components of a free Dirac 
spinor reduce to 2. These are normally the 
helicity ones, but a linear combinations 
of these corresponding to a given transverse polarization  
is more comfortable when working in impact parameter space 
and/or with given $\vec S_T$ 
(see \cite{AB_JPG1} for the technicalities of this choice, 
and \cite{GoldsteinMoravcsik82} for the properties of the 
scattering amplitudes in transverse-spin basis). 
I assume a transverse spin quantization 
axis $\hat x$ and 
split the UR quark  
wavefunction into two components corresponding to $s_x$ $=$ 
$\pm \hbar/2$, and build 2-component 
vectors: 
\begin{eqnarray} 
|\psi>\ \equiv\ \psi_+ |+>_x\ +\ \psi_- |->_x , \ \rightarrow\ 
\vec \psi(...)\ \equiv\ 
\left(
\begin{array}{cc}
\psi_+(...) \\
\psi_-(...) 
\end{array}
\right)
\label{eq:vec1} 
\end{eqnarray} 
where
(...) means spacetime or momentum variables. In this basis  
a single-particle amplitude $A$ must be expressed as a 
2x2 matrix operator. 


I name $q_1,q_2,...$ and $p_1,p_2,...$ the partons from hadrons 
``1'' and ``2'' respectively; $q_1$ and $p_1$ are the active 
quark and antiquark. $Q_\mu$ is the measured momentum 
of the Drell-Yan virtual photon, while $\vec q_T$ is the momentum 
exchanged between $q_1$ and $p_2,p_3,...$ in the 
rescattering event. The full series of terms appearing 
in the Drell-Yan cross section is  
listed in \cite{ArnoldMetzSchlegel09}. It follows the 
scheme 
\begin{equation}
\sigma\ \propto\ Im(fig.1 + fig.2 + ...)\ \equiv\ 
\sum d(x,x',Q_T,\{\theta\})\ =\ 
\end{equation}
\begin{equation}
=\ \sum \int d^2k_T d_1(x,\vec k_T)\cdot d_2(x',\vec Q_T - \vec k_T)
\label{eq:SF1}
\end{equation}
where the $d(x,x',...)$ structure functions 
are directly observable, while the tensor 
substructures $d_{1,2}(x,\vec k_T,..)$ 
(proportional to the TMDF) 
assume a factorization formalism and must be somehow reconstructed.  
$\{\theta\}$ is a set of dilepton angles.  


\subsection{Multiparton rescattering amplitude $A_N$}

For $x$ $<$ 1 GeV/$\sqrt{s}$ the 
wavelength of the quark $q_1$ 
from hadron ``1'' is not able to sample the longitudinal features 
of the set of spectator partons $p_2,p_3,p_4, ...$ coming from hadron 
``2'',  so $A$ may only affect 
the transverse degrees of 
freedom of 
$q_1$: 
\begin{eqnarray}
d(x,x',Q_T)\ =\  
<q_1|... \Big( 1\ +\ <p_2,p_3,...|A(\vec q_T)|p_2,p_3,...> \Big)....|q_1>
\nonumber
\\
\equiv\ <q_1|... \big( 1\ +\ A_N(\vec q_T) \big) ...|q_1>
\label{eq:rescattering}
\end{eqnarray}
where 
$A_N(\vec q_T)$ $\equiv$ $<p_2,p_3,..|A(\vec q_T)|p_2,p_3,..>$ is the 
first order sum/average of the rescattering amplitude over 
$N-1$ spectators, and ``1'' the no-rescattering term. 

$A_N$ is a 2x2 matrix 
acting on a $q_1$ wavefunction of the form \ref{eq:vec1}. 
In unpolarized and single-polarized Drell-Yan, 
the leading twist T-odd TMDF appear in $d_i(x,k_T)$ terms of 
the form (see \cite{Trento}) 
$d_{i,odd}(x,k_T)$ $=$ 
$h(x,\vec k_{i,T}) (\vec S_T \wedge \vec k_T)_z$
where 
$\vec S$ is the spin of one of the colliding hadrons (Sivers  
case) or of the active quark (Boer-Mulders case). 
The 
most general rescattering amplitude that is able to 
produce a term of this form 
(in a single-rescattering 
diagram, and without violating standard physical constraints 
including T-reversal) is: 
\begin{equation}
A_N(\vec q_T)\ \equiv\ A_{N, scalar} + A_{N, SO}\ =\  
f_N(q_T^2) +  g_N(q_T^2)
\bigg( \vec \sigma_T \wedge \vec q_T \bigg) 
\label{eq:scatt1}
\end{equation}
\begin{equation}
=\ 
\int d^2\vec b exp(i \vec q_T \cdot \vec b)
\Bigg\{ 
\hat f_N(b^2)\ +\ i 
\vec \sigma \wedge {\partial \over {\partial \vec b}}
\ \hat g_N(b^2)
\Bigg\} 
\label{eq:scatt2a}
\end{equation}
where 
``SO'' means ``spin-orbit'', 
(``spin-orbit'' and ``scalar'' in the sense defined in 
the Introduction: able or not to change $L$ by one unit). 
The momentum $\vec q_T$ exchanged in rescattering is a loop 
momentum, and 
must neither be confused with 
the active quark/antiquark momenta 
$\vec k_{i,T}$ nor with $\vec Q_T$ $=$ $\vec k_{1,T} + \vec k_{2,T}$ 
(see fig.2). 
The ``color thickness'' functions 
$f_N$ and $g_N$ derive from the averaging procedure 
$<p_2,p_3..|...|p_2,p_3..>$ in eq.\ref{eq:rescattering}. 
If $d(x,x',...)$ is organized in a factorized formalism as in 
eq.\ref{eq:SF1}, 
the modifications associated with fig.2 are attached to 
$d_2(x',\vec k_{2,T}$), that shows T-odd asymmetries  
$\propto$ 
$g_N(q_T^2)/[1\ +\ f_N(q_T^2)]$ $\approx$ $g_N(q_T^2)$ (the ``1'' comes 
from the ``1'' of eq. \ref{eq:rescattering}, i.e. from fig.1). 

Let 
both the scalar and the spin-orbit part of $A_N$ be 
the coherent sum of $single$ rescattering 
amplitudes, each one referring to boson exchange between $q_1$ and 
one of the 
$p_i$: 
\begin{eqnarray}
A_{N, scalar / SO}\  
=\ \nonumber \\
\sum_2^N c_i 
\int d^2b\ e^{i \vec q_T \vec b}\
\prod_2^N d^2b_j \ 
\hat O_{scalar , SO}\ 
P(q_T,|\vec b - \vec b_i|) 
\ |\psi(b_2,b_3,..)|^2
\label{eq:multiple1}
\end{eqnarray}
where 
$c_i$ are color charges and their sum must be 1. 
$\hat O_{scalar}$ $\equiv$ $\hat I$,   
$\hat O_{SO}$ $\equiv$ $\vec \sigma_T \wedge  
\partial / \partial \vec b$. The amplitude 
$P(q_T |\vec b - \vec b_i|)$ 
is the scalar interaction propagator in the 
transverse plane (that results after integrating over the longitudinal 
degrees of freedom). It has the general 
form 
\begin{equation}
P(q_T,|\vec b - \vec b_i|)\ =\ 
exp\Big(-i q_T |\vec b-\vec b_i|\Big)\ 
P'\Bigg({{|\vec b - \vec b_i|} \over {R_{int}}}\Bigg)\ 
\label{eq:propagator}
\end{equation}
where the strength factor $P'$ makes the individual quark-parton  
interaction effective within a distance $R_{int}$, 
presumably of hadronic size. I will assume $R_{int}$ to be 
$N-$independent, or at least weakly $N-$dependent w.r.t. $R_N$. 

The other relevant transverse scale parameter is  
$R_N$, the average transverse radius of the set of spectator 
partons, and for it I assume $R_N^2$ $=$ $(N-1)R_o^2$
 $\approx$ $NR_o^2$, where $R_o$ is a hadron-size parameter. 

Analytically, it is possible to study the two limiting cases 
$R_N$ $>>$ $R_{int}$ and $R_N$ $<<$ $R_{int}$, 
exploiting part of the standard 
procedure for describing scattering on a composite target 
in terms of charge and current form factors. 
I just outline the main steps. 

\subsection{case 1: The large$-N$ limit: $R_N$ $>>$ $R_{int}$} 

For $R_N$ $>>$ $R_{int}$, I may use assume a pointlike interaction 
range: 
\begin{equation}
P(q_T ,|\vec b - \vec b_i|)\ \rightarrow\ \delta(|\vec b - \vec b_i|).  
\label{eq:case1}
\end{equation}
\begin{eqnarray}
A_{N, scalar}\ 
\rightarrow\ \sum_2^N c_i 
\prod_2^N d^2b_j \ e^{i \vec q_T \vec b_i}
\ |\psi(b_2,b_3,..)|^2\ 
=\ F_c(q_T) 
\label{eq:short1}
\end{eqnarray}
where 
$F_c(q_T)$ is the color charge form-factor of the spectator parton 
set. 
 
For the spin-orbit case one uses the fact that in eq.\ref{eq:multiple1} 
$\partial/\partial \vec b$ 
only acts on 
$P(q_T,|\vec b - \vec b_i|)$, 
since all the other functions on the right 
of  $\partial/\partial \vec b$ depend on $\vec b_i$ but not on $\vec b$.  
For each 
$i$,  
\begin{equation}
{\partial \over {\partial \vec b}}\ P(q_T,|\vec b - \vec b_i|)\ K(\vec b_i)
\ =\ -\ 
K(\vec b_i)\ {\partial \over {\partial \vec b_i}}\ P(q_T,|\vec b - \vec b_i|),
\label{eq:reverse}
\end{equation}
where 
$K(b_i)$ summarizes all the terms in eq. \ref{eq:multiple1} that 
depend on $\vec b_i$ but not on $\vec b$. 
Now one may apply eq. \ref{eq:case1}, and 
integrate 
in $d \vec b$. One gets  
\begin{equation} 
A_{N, SO}(\vec q_T)\ =\ 
\sigma_T \wedge \vec q_T\ F_c(q_T)
\label{eq:short2}
\end{equation}

\subsection{Case 2: $R_N$ $<<$ $R_{int}$ 
and far away collisions: The small$-N$ cutoff.}

I examine the case where  
the spectators $p_i$ are close each other and far from $q_1$. 
Rescattering may take place if $R_N$ $<<$ $R_{int}$.  
Let me first start with the scalar term. 
\begin{equation}
exp\big(i \vec q_T \cdot \vec b\big)\ exp\big(-i q_T |\vec b - \vec b_i|)
\ \approx\ 
exp\big[i b q_T(\hat q_T \cdot \hat b - 1)\big] 
exp\big(i \vec b \cdot \vec b_i\big) 
\label{eq:approx1}
\end{equation}
\begin{equation} 
\rightarrow\ 
exp\big[i b q_T(\hat q_T \cdot \hat b - 1)\big] 
exp\big(i \vec q \cdot \vec b_i\big). 
\label{eq:case2}
\end{equation} 
The first member 
joins the two exponents from 
eqs. \ref{eq:multiple1} and \ref{eq:propagator}, eq. \ref{eq:approx1}
assumes $b$ $>>$ $b_i$ (I put the origin  
in between $p_2,..p_n$), and I 
use that in further integrals 
$exp[i b q_T (\hat q_T \cdot \hat b - 1)]$ will contribute  
for $\hat q_T \cdot \hat b$ $\approx$ 1. 
Eq. \ref{eq:multiple1} becomes
\begin{equation}
A_{N, scalar}(q_T)\ \approx\ 
P(q_T) 
\int \prod_2^N d^2b_j \ 
\Big(
\sum_2^N c_i e^{i \vec q_T \vec b_i}\
\Big) 
\ |\psi(b_2,b_3,..)|^2\ =
\end{equation}
\begin{equation}
=\ 
P(q_T)\ F_c(q_T)\ \approx\ P(q_T).
\label{eq:long1}
\end{equation}
where $P(q_T)$ is the amplitude for interacting with one pointlike 
constituent of charge 1. The last approximation reflects the fact 
that, because of the requirement $b$ $>>$ $b_i$,  
$P(q_T)$ is much more $q_T-$short-ranged than $F_c(q_T)$. 

For treating $A_{N, SO}$ one again uses eq. \ref{eq:reverse}. 
The final result is 
\begin{equation}
A_{N, SO}(\vec q_T)\ \approx\ 
P(q_T)\ F_c(q_T)\ \sigma_T \wedge \vec q_T 
\ \approx 
P(q_T)\ \sigma_T \wedge \vec q_T 
\label{eq:long2}
\end{equation}
So, 
we have a cutoff. For small enough $N$ to have $R_N$ $\leq$ $R_{int}$, 
far away rescattering is possible and likely. 
Then spectators appear as 
a unique pointlike constituent. This case corresponds to a proper 
use of the diquark approximation\cite{JakobMuldersRodriguez97}. 
For large $N$ and $R_N$ $>$ $R_{int}$ 
the increased  sensitivity 
to the structure of the spectator ``cloud'' is 
expressed by eqs. \ref{eq:short1} and \ref{eq:short2}. 
This allows for defining the lower cutoff for the ``large $N$'' 
condition used in this work. It means that the 
equation $\sqrt{N}R_o$ $>$ $R_{int}$ is satisfied. 

\subsection{Dependence of the observable asymmetries on $N$}

Comparing the form-factor expressions 
with eq.\ref{eq:scatt2a}, we see that 
\begin{eqnarray}
\hat f_N(b^2)\ \propto\ F_c(q_T),\ 
\hat g_N(b^2)\ \propto\ F_c(q_T). 
\label{eq:multiple7}
\end{eqnarray}
At 
2nd order in $q_TR_N$ 
\begin{equation}
F_c(q_T)\ \approx\ exp(- q_T^2 R_N^2 / 2)\ 
=\ exp(- N q_T^2 R_o^2 / 2)\ 
\label{eq:formfactor1}
\end{equation}
The 
scalar term may be large for $q_T$ $\approx$ 0. In the spin-orbit 
case this possibility is forbidden 
since 
\begin{equation}
A_{N, SO}(\vec q_T)\ \propto\ 
\vec \sigma_T \wedge \vec q_T\ F_c(q_T)\ \propto\ 
\vec \sigma_T \wedge \vec q_T\ exp(- N q_T^2 R_o^2 / 2). 
\label{eq:formfactor2}
\end{equation}
$A_N(\vec q_T)$ is a matrix on spin states. For a given spin, it 
is odd in $\vec q_T$, negligible 
for $|q_T|$ $>$ $2/\sqrt{N}R_o$, and roughly linear up to its 
peak at $q_T$ $\approx$ 
$1/\sqrt{N}R_o$: 
\begin{equation}
Max|A_{N, SO}(\vec q_T)|\ \approx\ A_{N, SO}(q_T = 1/\sqrt{N}R_o)\ 
\propto\ 1/\sqrt{N}.
\label{eq:formfactor3}
\end{equation}
We 
get observable effects from the 
convolution
\begin{equation} 
A_N(\vec Q_T)\ \equiv\ 
\int d^2 q_T\  G(\vec Q_T - \vec q_T)\ A_N(\vec q_T) 
\label{eq:fulldistribution1}
\end{equation}
where 
$\vec Q_T$ $\equiv$ $\vec k_{T,1}$ $+$ $\vec k_{T,2}$ $+$ $\vec q_T$ 
$\equiv$ $\vec k_T$ $+$ $\vec q_T$,  
and $G(\vec k_T)$ is a distribution for the quark/antiquark combined 
``intrinsic'' momentum, possibly including real-gluon radiation 
effects. 

In the ``quasi-collinear'' limit 
$\vec k_T$ is negligible, $\vec Q_T$ $\approx$ $\vec q_T$, 
$A_N(\vec Q_T)$ $\approx$ $A_N(\vec q_T)$.   
$A_N(Q_y)-A_N(-Q_y)$ $\approx$ 
$A_N(q_y)-A_N(-q_y)$ $=$ 
$A_{N,SO}(q_T)$ $\sim$ $1/\sqrt{N}$. 

However, in Drell-Yan $k_T$ receives strong 
contributions from hard real gluon radiation 
and $<k_T^2>$ overcomes the soft hadronic momentum scales 
(see \cite{JQVY06} and \cite{matches08} for the behavior of T-odd 
TMDF at large $Q_T$). For large $N$ the average $q_T$ is $O(1/\sqrt{N})$ 
and it seems likely to have $k_T$ $>>$ $q_T$. Using eq. 
\ref{eq:formfactor3}, and assuming a gaussian form for 
$F(Q_T)$,  
\begin{eqnarray}
A_N(Q_y)\ -\ A_N(-Q_y)\ =\ 
\int d^2 q_T\  G(\vec Q_T - \vec q_T)\ A_{N,SO}(\vec q_T)\ 
\sim\ \nonumber\\ \sim\ 
\Bigg\{ \big[ G( Q_T + 1/\sqrt{N}R_o )\big]\ -\ 
\big[ G(Q_T - 1/\sqrt{N}R_o) \big] \Bigg\}\ 
\Bigg[\int_0^{2/\sqrt{N}R_o}A_{N,SO}(q_y)dq_y\ \Bigg]\  
\nonumber
\end{eqnarray}
\begin{equation}
\sim\ 
(2/\sqrt{N}R_o)^2 
\cdot  
(2/\sqrt{N}R_o)
{\partial \over {\partial Q_T}} F(Q_T)\ \propto\ 
Q_T G(Q_T) (1/\sqrt{N})^3
\label{eq:finalestimate1}
\end{equation}

\section{Conclusions and discussion}

Summarizing, if the hypotheses of this work are valid 
(small $x$ and $x'$, one hadron in a high-order Fok state 
with $N$ partons) 
the observable T-odd effects of the 
rescattering amplitude in Drell-Yan 
are weakened by a factor ranging from 
$1/N^{1/2}$ ($Q_T$ $<<$ 1 GeV/c) to $1/N^{3/2}$ 
($Q_T$ $\sim$ 1 GeV/c). 

The key assumption of this work is that the diagrams describing 
the interactions between the active (anti)quark from 
hadron ``1'' and each of the spectator partons from ``2'' sum 
coherently, weighted by coefficients that are linear in the 
spectator color charges. This is obvious for 
short-distance 1-gluon exchange, less obvious for interactions 
that may in principle involve large transverse distances (although 
the main result of this work has been obtained in the approximation 
of short-ranged rescattering, see eq.\ref{eq:case1}). 

To stress the role of color coherence in the previous 
results, let me consider an alternative small$-x$ model, 
where I obtain a multiparton state by a chain 
of quasi-real and well-separated mesons, each in a pure valence 
state, instead of a chain of partons. 
This is a large$-N$ extension of 
the pion-pole limit of the pion-cloud model \cite{PasquiniBoffi06}. 
The proton-proton Drell-Yan cross section 
would be the convolution between the probability of finding a 
$\pi^-$ in a given place and the valence cross section of 
$\pi^--$proton Drell-Yan. Intuition suggests that if any 
suppression is present, in this model it is not 
T-odd-selective. This extreme picture takes into account one only 
(``singlet $\otimes$ singlet $\otimes$ .... $\otimes$ triplet'')
among all the possible color space decompositions of the $N-$parton 
set. So it excludes by default amplitude cancellations with other 
possible color configurations. The multimeson picture is unlikely for 
large $N$ (the mesons would be highly virtual, overlap and lose 
individuality), but 
it would be interesting to see how much color cancellations do/do not 
suppress T-odd effects in the 5-parton case of the 
model \cite{PasquiniBoffi06}. 

An important discussion hint is the widespread use, to calculate 
T-odd observables, of the diquark 
spectator model\cite{JakobMuldersRodriguez97} (see e.g. 
\cite{BrodskyHwangSchmidt02,BoerBrodskyHwang03, 
GambergGoldsteinOganessyan03,BacchettaSchaeferYang04,
GoekeMeissnerMetzSchlegel06,BacchettaContiRadici08}) 
or of models that only include the lowest-Fok state 
(e.g. \cite{Yuan03,LuMa04,Courtoy08,AB_JPG1,Vento09,PasquiniYuan10}). 
Although large$-N$ states are relevant at small$-x$, 
my analysis says that the contribution from these states to 
T-odd effects is much smaller than in the T-even case. 
So, valence-model predictions could  be better than 
presently imagined.  

The risk however is the ``opposite'' error: a model that 
effectively includes 
sea properties in a valence distribution (e.g. a model that  
produces a valence quark distribution with typical sea magnitude 
at small $x$) will overestimate T-odd effects at small $x$. Indeed, 
it will make many wee partons rescatter like a single pointlike 
one. In this 
case, I would recommend the introduction of a gluon-spectator 
form factor to take finite spectator size into account.




\end{document}